\documentclass[aps,prl,twocolumn,superscriptaddress,showpacs,preprintnumbers,amsmath,amssymb]{revtex4-1}
\usepackage{graphicx}
\usepackage{dcolumn}
\RequirePackage{xspace}
\usepackage{ulem}
\usepackage{xcolor}

\def\PHI  {\ensuremath{\phi}\xspace}
\def\ETAc  {\ensuremath{\eta_c}\xspace}

\def\qqbar {\ensuremath{q\overline q}\xspace}
\def\sbar  {\ensuremath{\overline s}\xspace}
\def\EP   {\ensuremath{e^+}\xspace}
\def\EM   {\ensuremath{e^-}\xspace}

\def\pip  {\ensuremath{\pi^+}\xspace}
\def\pim  {\ensuremath{\pi^-}\xspace}
\def\Kbar  {\kern 0.2em\overline{\kern -0.2em K}{}\xspace}

\def\Kz   {\ensuremath{K^0}\xspace}
\def\Kp   {\ensuremath{K^+}\xspace}
\def\Km   {\ensuremath{K^-}\xspace}
\def\kpm  {\ensuremath{K^{\pm}}\xspace}
\def\KS   {\ensuremath{K^0_{\scriptscriptstyle S}}\xspace} 
\def\Dbar  {\kern 0.2em\overline{\kern -0.2em D}{}\xspace}

\def\Dz   {\ensuremath{D^0}\xspace}
\def\Dzb  {\ensuremath{\Dbar^0}\xspace}
\def\Dsp  {\ensuremath{D^+}_{\scriptscriptstyle s}\xspace}
\def\Dstarp {\ensuremath{D^{*+}}\xspace}
\def\Bbar  {\kern 0.18em\overline{\kern -0.18em B}{}\xspace}
\def\BB   {\ensuremath{B\Bbar}\xspace} 
\def\Bz   {\ensuremath{B^0}\xspace}
\def\Bu   {\ensuremath{B^+}\xspace}
\def\Bub  {\ensuremath{B^-}\xspace}
\def\Bp   {\ensuremath{\Bu}\xspace}
\def\Bm   {\ensuremath{\Bub}\xspace}
\def\Bpm  {\ensuremath{B^{\pm}}\xspace}
\mathchardef\Upsilon="7107
\def\Y#1S{\ensuremath{\Upsilon{(#1S)}}\xspace}
\def\mbc  {\mbox{$M_{\rm bc}$}\xspace}
\def\DeltaE {\mbox{$\Delta E$}\xspace}
\def\cm  {\ensuremath{{\rm \,cm}}\xspace}
\def\invfb{\ensuremath{\mbox{\,fb}^{-1}}\xspace}
\def\CP {\ensuremath{C\!P}\xspace}
\def\ACP{{\ensuremath{\mathcal{A}_{\CP}}\xspace}}
\def\ACPj{{\ensuremath{\mathcal{A}_{\CP,j}}\xspace}}
\def\etal {{\it et~al.}}
\def\nb  {\ensuremath{O_{\rm N\!N}}\xspace}
\def\nbprim{\ensuremath{O'_{\rm N\!N}}\xspace}
\def\nbmin {\ensuremath{O_{\rm N\!N,{\rm min}}}\xspace}
\def\nbmax {\ensuremath{O_{\rm N\!N,{\rm max}}}\xspace}
\newcommand{\stat}{\ensuremath{\mathrm{(stat)}}\xspace}
\newcommand{\syst}{\ensuremath{\mathrm{(syst)}}\xspace}
\newcommand{\gev}{\ensuremath{\mathrm{\,Ge\kern -0.1em V}}\xspace}
\newcommand{\mev}{\ensuremath{\mathrm{\,Me\kern -0.1em V}}\xspace}
\newcommand{\gevcc}{\ensuremath{{\mathrm{\,Ge\kern -0.1em V\!/}c^2}}\xspace}
\newcommand{\mevcc}{\ensuremath{{\mathrm{\,Me\kern -0.1em V\!/}c^2}}\xspace}

\begin{document}

\preprint{\vbox{ \hbox{  }
			\hbox{Belle Preprint {\it 2020-20}}
			\hbox{KEK Preprint {\it 2020-37}}
}}

\title{\quad\\[1.0cm] Measurement of Branching Fraction and Search for {\boldmath $\CP$} Violation in {\boldmath $B\to\PHI\PHI K$}}

\noaffiliation
\affiliation{University of the Basque Country UPV/EHU, 48080 Bilbao}
\affiliation{University of Bonn, 53115 Bonn}
\affiliation{Brookhaven National Laboratory, Upton, New York 11973}
\affiliation{Budker Institute of Nuclear Physics SB RAS, Novosibirsk 630090}
\affiliation{Faculty of Mathematics and Physics, Charles University, 121 16 Prague}
\affiliation{Chonnam National University, Gwangju 61186}
\affiliation{University of Cincinnati, Cincinnati, Ohio 45221}
\affiliation{Deutsches Elektronen--Synchrotron, 22607 Hamburg}
\affiliation{University of Florida, Gainesville, Florida 32611}
\affiliation{Key Laboratory of Nuclear Physics and Ion-beam Application (MOE) and Institute of Modern Physics, Fudan University, Shanghai 200443}
\affiliation{SOKENDAI (The Graduate University for Advanced Studies), Hayama 240-0193}
\affiliation{Gyeongsang National University, Jinju 52828}
\affiliation{Department of Physics and Institute of Natural Sciences, Hanyang University, Seoul 04763}
\affiliation{University of Hawaii, Honolulu, Hawaii 96822}
\affiliation{High Energy Accelerator Research Organization (KEK), Tsukuba 305-0801}
\affiliation{J-PARC Branch, KEK Theory Center, High Energy Accelerator Research Organization (KEK), Tsukuba 305-0801}
\affiliation{Higher School of Economics (HSE), Moscow 101000}
\affiliation{Forschungszentrum J\"{u}lich, 52425 J\"{u}lich}
\affiliation{IKERBASQUE, Basque Foundation for Science, 48013 Bilbao}
\affiliation{Indian Institute of Science Education and Research Mohali, SAS Nagar, 140306}
\affiliation{Indian Institute of Technology Bhubaneswar, Satya Nagar 751007}
\affiliation{Indian Institute of Technology Hyderabad, Telangana 502285}
\affiliation{Indian Institute of Technology Madras, Chennai 600036}
\affiliation{Indiana University, Bloomington, Indiana 47408}
\affiliation{Institute of High Energy Physics, Chinese Academy of Sciences, Beijing 100049}
\affiliation{Institute of High Energy Physics, Vienna 1050}
\affiliation{Institute for High Energy Physics, Protvino 142281}
\affiliation{INFN - Sezione di Napoli, 80126 Napoli}
\affiliation{INFN - Sezione di Torino, 10125 Torino}
\affiliation{Advanced Science Research Center, Japan Atomic Energy Agency, Naka 319-1195}
\affiliation{J. Stefan Institute, 1000 Ljubljana}
\affiliation{Institut f\"ur Experimentelle Teilchenphysik, Karlsruher Institut f\"ur Technologie, 76131 Karlsruhe}
\affiliation{Kennesaw State University, Kennesaw, Georgia 30144}
\affiliation{Department of Physics, Faculty of Science, King Abdulaziz University, Jeddah 21589}
\affiliation{Kitasato University, Sagamihara 252-0373}
\affiliation{Korea Institute of Science and Technology Information, Daejeon 34141}
\affiliation{Korea University, Seoul 02841}
\affiliation{Kyungpook National University, Daegu 41566}
\affiliation{Universit\'{e} Paris-Saclay, CNRS/IN2P3, IJCLab, 91405 Orsay}
\affiliation{P.N. Lebedev Physical Institute of the Russian Academy of Sciences, Moscow 119991}
\affiliation{Faculty of Mathematics and Physics, University of Ljubljana, 1000 Ljubljana}
\affiliation{Luther College, Decorah, Iowa 52101}
\affiliation{Malaviya National Institute of Technology Jaipur, Jaipur 302017}
\affiliation{University of Maribor, 2000 Maribor}
\affiliation{Max-Planck-Institut f\"ur Physik, 80805 M\"unchen}
\affiliation{School of Physics, University of Melbourne, Victoria 3010}
\affiliation{University of Miyazaki, Miyazaki 889-2192}
\affiliation{Graduate School of Science, Nagoya University, Nagoya 464-8602}
\affiliation{Universit\`{a} di Napoli Federico II, 80126 Napoli}
\affiliation{Nara Women's University, Nara 630-8506}
\affiliation{National Central University, Chung-li 32054}
\affiliation{National United University, Miao Li 36003}
\affiliation{Department of Physics, National Taiwan University, Taipei 10617}
\affiliation{H. Niewodniczanski Institute of Nuclear Physics, Krakow 31-342}
\affiliation{Niigata University, Niigata 950-2181}
\affiliation{Novosibirsk State University, Novosibirsk 630090}
\affiliation{Osaka City University, Osaka 558-8585}
\affiliation{Pacific Northwest National Laboratory, Richland, Washington 99352}
\affiliation{Peking University, Beijing 100871}
\affiliation{University of Pittsburgh, Pittsburgh, Pennsylvania 15260}
\affiliation{Punjab Agricultural University, Ludhiana 141004}
\affiliation{Research Center for Nuclear Physics, Osaka University, Osaka 567-0047}
\affiliation{Meson Science Laboratory, Cluster for Pioneering Research, RIKEN, Saitama 351-0198}
\affiliation{Department of Modern Physics and State Key Laboratory of Particle Detection and Electronics, University of Science and Technology of China, Hefei 230026}
\affiliation{Seoul National University, Seoul 08826}
\affiliation{Showa Pharmaceutical University, Tokyo 194-8543}
\affiliation{Soochow University, Suzhou 215006}
\affiliation{Soongsil University, Seoul 06978}
\affiliation{Sungkyunkwan University, Suwon 16419}
\affiliation{School of Physics, University of Sydney, New South Wales 2006}
\affiliation{Department of Physics, Faculty of Science, University of Tabuk, Tabuk 71451}
\affiliation{Tata Institute of Fundamental Research, Mumbai 400005}
\affiliation{School of Physics and Astronomy, Tel Aviv University, Tel Aviv 69978}
\affiliation{Toho University, Funabashi 274-8510}
\affiliation{Department of Physics, Tohoku University, Sendai 980-8578}
\affiliation{Earthquake Research Institute, University of Tokyo, Tokyo 113-0032}
\affiliation{Department of Physics, University of Tokyo, Tokyo 113-0033}
\affiliation{Tokyo Institute of Technology, Tokyo 152-8550}
\affiliation{Tokyo Metropolitan University, Tokyo 192-0397}
\affiliation{Utkal University, Bhubaneswar 751004}
\affiliation{Virginia Polytechnic Institute and State University, Blacksburg, Virginia 24061}
\affiliation{Wayne State University, Detroit, Michigan 48202}
\affiliation{Yamagata University, Yamagata 990-8560}
\affiliation{Yonsei University, Seoul 03722}
  \author{S.~Mohanty}\affiliation{Tata Institute of Fundamental Research, Mumbai 400005}\affiliation{Utkal University, Bhubaneswar 751004} 
  \author{A.~B.~Kaliyar}\affiliation{Tata Institute of Fundamental Research, Mumbai 400005} 
  \author{V.~Gaur}\affiliation{Virginia Polytechnic Institute and State University, Blacksburg, Virginia 24061} 
  \author{G.~B.~Mohanty}\affiliation{Tata Institute of Fundamental Research, Mumbai 400005} 
  \author{I.~Adachi}\affiliation{High Energy Accelerator Research Organization (KEK), Tsukuba 305-0801}\affiliation{SOKENDAI (The Graduate University for Advanced Studies), Hayama 240-0193} 
  \author{K.~Adamczyk}\affiliation{H. Niewodniczanski Institute of Nuclear Physics, Krakow 31-342} 
  \author{H.~Aihara}\affiliation{Department of Physics, University of Tokyo, Tokyo 113-0033} 
  \author{S.~Al~Said}\affiliation{Department of Physics, Faculty of Science, University of Tabuk, Tabuk 71451}\affiliation{Department of Physics, Faculty of Science, King Abdulaziz University, Jeddah 21589} 
  \author{D.~M.~Asner}\affiliation{Brookhaven National Laboratory, Upton, New York 11973} 
  \author{H.~Atmacan}\affiliation{University of Cincinnati, Cincinnati, Ohio 45221} 
% ...
\collaboration{The Belle Collaboration}
\noaffiliation

\begin{abstract}
We report the measurement of branching fractions and $\CP$-violation asymmetries in $B\to\PHI\PHI K$ 
decays based on a $711\invfb$ data sample containing $772\times 10^6$ $\BB$ events. The data were
recorded at the $\Y4S$ resonance with the Belle detector at the KEKB asymmetric-energy $\EP\EM$ collider.
For $\Bp\to\PHI\PHI\Kp$, the branching fraction and $\CP$-violation asymmetry measured below the
$\ETAc$ threshold ($m_{\PHI\PHI}<2.85\gevcc$) are $[3.43^{\,+\,0.48}_{\,-\,0.46}\stat\pm 0.22\syst]
\times10^{-6}$ and $-0.02\pm0.11\stat\pm0.01\syst$, respectively. Similarly, the branching fraction
obtained for $\Bz\to\PHI\PHI\Kz$ below the $\ETAc$ threshold is $[3.02^{\,+\,0.75}_{\,-\,0.66}
\stat\pm \,0.20\syst]\times10^{-6}$. We also measure the $\CP$-violation asymmetry for
$\Bp\to\PHI\PHI\Kp$ within the $\ETAc$ region ($m_{\PHI\PHI}\in [2.94,3.02]\gevcc$) to be
$+0.12\pm0.12\stat\pm0.01\syst$. 
\end{abstract}

\pacs{13.25.Hw, 14.40.Nd}

\maketitle

\tighten

{\renewcommand{\thefootnote}{\fnsymbol{footnote}}}
\setcounter{footnote}{0}

\begin{figure}[!htb]
\begin{center}$
\begin{array}{cc}
\includegraphics[width=.24\textwidth]{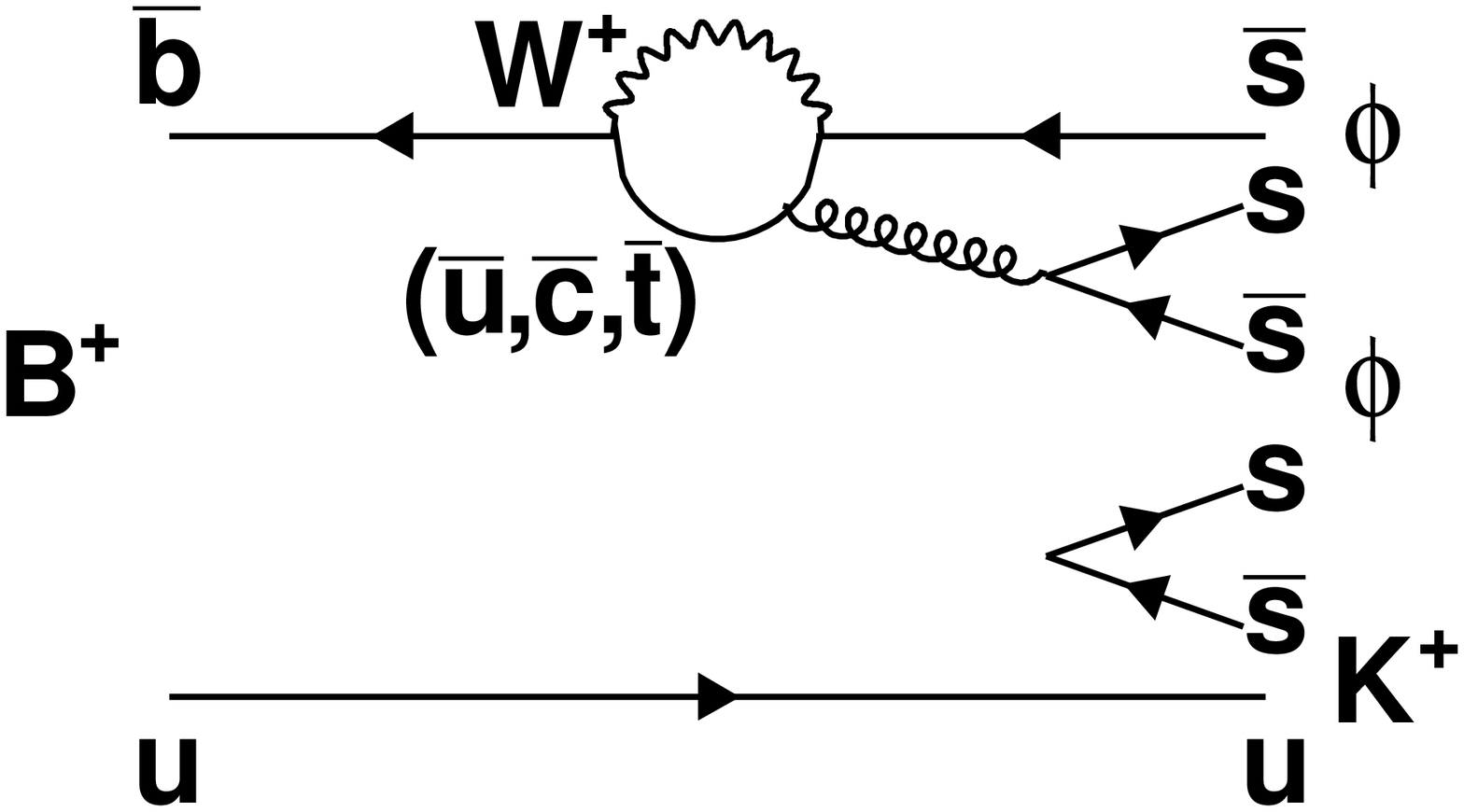} &
\includegraphics[width=.24\textwidth]{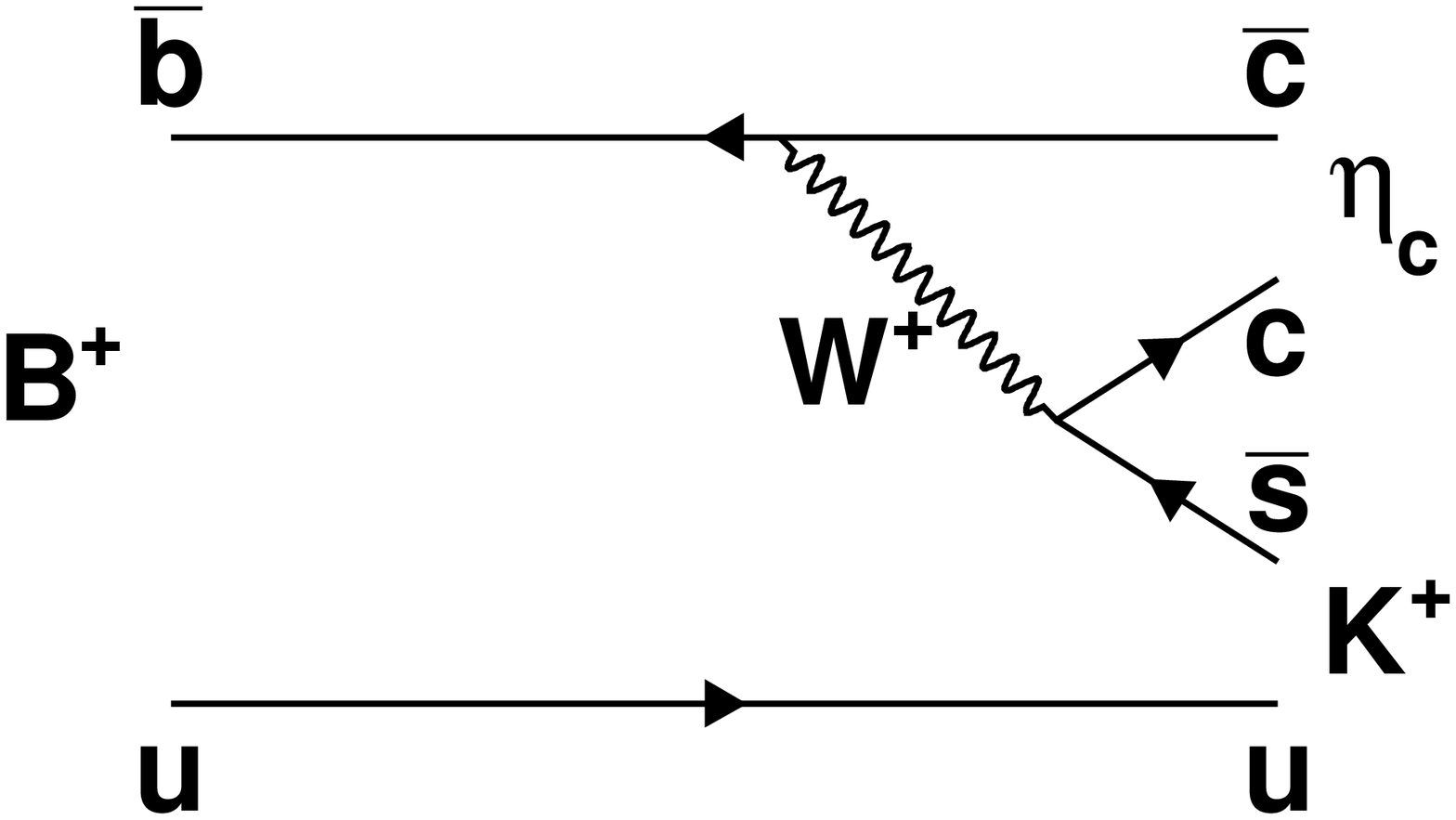} \\
\end{array}$
\end{center}
\caption{Dominant Feynman diagrams that contribute to the decays (left) $\Bp\to\PHI\PHI\Kp$ and
(right) $\Bp\to\ETAc\Kp$. Replacement of the spectator $u$ quark with a $d$ quark will lead to the
corresponding diagrams for $\Bz\to\PHI\PHI\Kz$ and $\Bz\to\ETAc\Kz$.}
\label{fig:Feyn_diag}
\end{figure}

$B$-meson decays to three-body $\PHI\PHI K$ final states proceed via a $b\to s\sbar s$ loop
(penguin) transition, which requires the creation of an additional $s\sbar$ pair. The same
final state can also originate from the tree-level process $B\to\ETAc(\to\PHI\PHI)K$.
Figure~\ref{fig:Feyn_diag} shows the dominant Feynman diagrams that contribute to these
decays. The interference between penguin and tree amplitudes is maximal when the $\PHI\PHI$
invariant mass lies close to the $\ETAc$ mass $(m_{\phi\phi}\in[2.94,3.02]\gevcc)$. No $\CP$
violation is expected from this interference, as the relative weak phase between the two
amplitudes is ${\rm arg}(V_{tb}V_{ts}^{*}/V_{cb}V_{cs}^{*})\approx 0$, where $V_{ij}$ denote
CKM matrix elements~\cite{Ref:ckm}. A potential new physics (NP) contribution to the loop,
however, can introduce a nonzero $\CP$-violating phase. In particular, the $\CP$ asymmetry can
be as large as $40\%$ in the presence of NP~\cite{Ref:Hazumi}. Thus, an observation of large $\CP$
violation in $B\to\PHI\PHI K$ would indicate the presence of physics beyond the Standard Model.
In addition to being an NP probe, the decay is sensitive to the possible production of a glueball
candidate near $2.3\,\gevcc$ that can subsequently decay to $\PHI\PHI$~\cite{Ref:Chua}. We can
also search for a structure at $2.35\,\gevcc$ observed in the $m_{\phi\phi}$ distribution in
two-photon collisions~\cite{Ref:Liu2012eb} and dubbed the $X(2350)$.

Based on a $78\invfb$ data sample, Belle reported the first evidence for the decay with a branching fraction 
${\cal B}(\Bp\to\PHI\PHI\Kp)=[2.6^{\,+\,1.1}_{\,-\,0.9}\stat\pm0.3\syst]\times 10^{-6}$~\cite{Ref:Huang} below
the $\ETAc$ threshold $(m_{\PHI\PHI}<2.85\gevcc)$~\cite{Ref:Inclusion}. The result was consistent with the
corresponding theory prediction, which lies in the range $(1.3$--$4.2)\times 10^{-6}$~\cite{Ref:theory1,Ref:theory2}.
The BaBar experiment performed a measurement of this decay using their full dataset of $464\times 10^{6}$ $\BB$
events~\cite{Ref:Atmacan}. The branching fraction obtained with the same $m_{\PHI\PHI}$ requirement was
${\cal B}(\Bp\to\PHI\PHI\Kp)=(5.6\pm 0.5\pm 0.3)\times 10^{-6}$, about three standard deviations above Belle's
result and larger than theoretical estimates. The $\Bz\to\PHI\PHI\Kz$ channel was observed with a branching
fraction of $(4.5 \pm 0.8 \pm 0.3)\times 10^{-6}$. BaBar also reported $\CP$ asymmetries for charged $B$
decays as $-0.10\pm 0.08\pm 0.02$ below the $\ETAc$ threshold and $+0.09\pm0.10\pm0.02$ within the $\ETAc$
region.

In this paper, we update our earlier result~\cite{Ref:Huang} with a significantly larger data sample containing
$772\times 10^6$ $\BB$ events. The data were collected at the $\Y4S$ resonance with the Belle detector~\cite{Belle}
at the KEKB asymmetric-energy $\EP\EM$ collider~\cite{KEKB}. The subdetectors relevant for our study are
a silicon vertex detector (SVD), a central drift chamber (CDC), an array of aerogel threshold Cherenkov
counters (ACC), and time-of-flight scintillation counters (TOF). All these are located inside
a $1.5$\,T axial magnetic field.

To reconstruct $\Bp\to\PHI\PHI\Kp$ and $\Bz\to\PHI\PHI\Kz$ decay candidates, we combine a pair of $\PHI$
mesons with a charged kaon and $\KS$, respectively. All charged tracks except for those from the $\KS$ must
have a distance of closest approach with respect to the interaction point (IP) of less than $0.2\cm$ in the
transverse $r$--$\phi$ plane, and less than $5.0\cm$ along the $z$ axis. The $z$ axis is defined as the direction
opposite that of the $e^+$ beam. We identify charged kaons based on a likelihood ratio ${\cal R}_{K/\pi}
={{\cal L}_K}/({\cal L}_K+{\cal L}_\pi)$, where ${\cal L}_K$ and ${\cal L}_\pi$ denote the individual
likelihood for kaons and pions, respectively. These are calculated using specific ionization in the CDC
and information from the ACC and the TOF. A requirement ${\cal R}_{K/\pi}>0.6$ is applied to select kaon
candidates. The kaon identification efficiency, averaged over the momentum range, is $90\%$, with a pion
misidentification rate of about $10\%$.

We reconstruct the $\PHI$ candidates from pairs of oppositely charged kaons with an invariant mass in the range
$1.00$--$1.04\gevcc$, corresponding to $\pm 5\sigma$ ($\sigma$ is the width of the mass distribution) around the
nominal $\PHI$ mass~\cite{PDG}. This is referred to as the $M_{KK}$ signal region in the following discussion.
The $\KS$ candidates are reconstructed from two oppositely charged tracks, assumed to be pions, and are
further required to satisfy a criterion on the output of a neural network (NN) algorithm~\cite{neurobayes}.
The algorithm uses the following input variables: the $\KS$ momentum in the lab frame; the distance of closest
approach along the $z$ axis between the two pion tracks; the flight length in the $r$--$\phi$ plane; the angle
between the $\KS$ momentum and the vector joining the IP to the $\KS$ decay vertex; the angle between the $\KS$
momentum in the lab frame and the pion momentum in the $\KS$ rest frame; the distances of closest approach in
the $r$--$\phi$ plane between the IP and the two pion tracks; the number of CDC hits for each pion track; and
the presence or absence of SVD hits for each pion track. We require that the invariant mass lie between $491\mevcc$
and $504\mevcc$, which corresponds to a $\pm 3\sigma$ window in resolution around the nominal $\KS$ mass~\cite{PDG}. 

$B$-meson candidates are identified with two kinematic variables: the beam-energy-constrained mass $\mbc
\equiv\sqrt{E^{2}_{\rm b}/c^{4}-\left\lvert\sum_{i}\vec{p}_{i}/c\right\rvert^{2}}$, and the energy difference
$\DeltaE\equiv\sum_{i}E_{i}-E_{\rm b}$, where $E_{\rm b}$ is the beam energy, and $\vec{p}_{i}$ and $E_{i}$
are the momentum and energy, respectively, of the $i$-th decay product of the $B$ candidate. All these
quantities are evaluated in the $\EP\EM$ center-of-mass (CM) frame. We perform a fit for each $B$
candidate, constraining its decay products to originate from a common vertex. Candidate events with
$\mbc\in[5.230,5.289]\gevcc$ and $\lvert\DeltaE\rvert <0.1\gev$ are retained for further study.
The $\mbc$ requirement corresponds to approximately ($-16\sigma$,\,$+3\sigma$) in resolution around the nominal
$B$ mass~\cite{PDG}, and the $\DeltaE$ requirement denotes a $\pm 10\sigma$ window around zero. We apply
such loose requirements on $\mbc$ and $\DeltaE$ as these are used in a maximum-likelihood fit to obtain
the signal yield (described later). We define a signal region as $\mbc\in[5.272,5.289]\gevcc$ and
$\lvert\DeltaE\rvert<0.05\gev$.

After application of the above selection criteria, the average number of $B$ candidates found per event selected
in data are $1.7$ and $1.6$ for $\Bp\to\PHI\PHI\Kp$ and $\Bz\to\PHI\PHI\Kz$, respectively. In the case of multiple
$B$ candidates, we choose the candidate with the lowest $\chi^{2}$ value for the aforementioned $B$-vertex fit.
From Monte Carlo (MC) simulation the best candidate selection method is found to have an efficiency of $68\%$
($65\%$) to correctly identify the $B$-meson candidate in $\Bp\to\PHI\PHI\Kp$ ($\Bz\to\PHI\PHI\Kz$) decays. In only
about $6\%$ of the total signal events, the $B$ candidate is misreconstructed due to swapping of kaons between the
two $\PHI$ candidates, or of one daughter track with that from the rest of the event. Such misreconstructed events
are treated as a part of the signal.

The dominant background is from the $\EP\EM\to\qqbar$ ($q=u,d,s,c$) continuum process. To suppress this background,
observables based on event topology are used. The event shape in the CM frame is expected to be spherical for
$\BB$ events and jet-like for continuum events. We use an NN~\cite{neurobayes} to combine the following six 
variables: a Fisher discriminant formed out of $16$ modified Fox-Wolfram moments~\cite{KSFW}; the cosine of the
angle between the $B$ momentum and the $z$ axis; the cosine of the angle between the $B$ thrust axis~\cite{thrust}
and the $z$ axis; the cosine of the angle between the thrust axis of the $B$ candidate and that of the rest of the
event; the ratio of the second- to the zeroth-order Fox-Wolfram moments (all quantities are calculated in the CM
frame); and the vertex separation along the $z$ axis between the $B$ candidate and the remaining tracks. The NN
training and validation are performed with signal and $\qqbar$ MC simulated events. The signal sample is generated
with the \textsc{EvtGen} program~\cite{evtgen}, assuming a uniform distribution over the three-body phase space of
the final state.

The neural network output ($\nb$) ranges between $-1.0$ and $1.0$, where events near $-1.0$ ($1.0$) are more
continuum- (signal-) like. We apply a loose criterion $\nb>-0.5$ to reduce the continuum background. The
relative signal efficiency loss due to this requirement is about $6\%$ ($3\%$) for $\Bp\to\PHI\PHI\Kp$
($\Bz\to\PHI\PHI\Kz$) decays, whereas the fraction of continuum events rejected is $76\%$ ($66\%$). As the remainder
of the $\nb$ distribution strongly peaks near $1.0$ for signal, it is difficult to model with an analytic
function. However, the transformed variable
\begin{eqnarray}
\nbprim=\log\left[\frac{\nb-\nbmin}{\nbmax-\nb}\right], 
\end{eqnarray}
where $\nbmin=-0.5$ and $\nbmax\simeq 1.0$, has a Gaussian-like distribution that is easier to model.
Thus, we use this transformed variable in our signal fit.

Backgrounds due to $B$ decays, mediated by the dominant $b\to c$ transition, are studied
with MC samples of such decays. For both $\Bp\to\PHI\PHI\Kp$ and $\Bz\to
\PHI\PHI\Kz$ channels, the $\mbc$ and $\DeltaE$ distributions are found to peak in the
signal region. To investigate the source of these contributions, we inspect the $m_{\PHI\PHI}$
distribution, which displays several peaks corresponding to the $\ETAc$ and other
charmonium resonances. To suppress these peaking backgrounds, we exclude candidates
for which the $m_{\PHI\PHI}$ value is greater than $2.85\gevcc$. This requirement also
allows us to compare our results with the earlier ones from Belle~\cite{Ref:Huang}
and BaBar~\cite{Ref:Atmacan}. We calculate the detection efficiencies for
candidate events below the $\ETAc$ threshold to be $12.4\%$ and $12.0\%$ for $\Bp\to
\PHI\PHI\Kp$ and $\Bz\to\PHI\PHI\Kz$, respectively.

Charmless backgrounds that do not produce only kaons in the final state may still
contribute to the $\mbc$--$\DeltaE$ signal region when a final-state particle is
misidentified. These are studied with a $\BB$ MC sample in which one of the $B$
mesons decays via $b\to u,d,s$ transitions with known or estimated branching
fractions~\cite{PDG}. Only $40$ events survive from an MC sample equivalent to $50$
times the size of the data sample. This small component is combined with the events
surviving from $b\to c$ transitions to form an overall $\BB$ background component.
In addition to this $\BB$ background that does not peak in $\mbc$ or $\DeltaE$,
we can have contributions from $B\to\phi KKK$ and $B\to KKKKK$ decays (described
later), which have the same final-state particles as the signal.

The signal yield is obtained with an unbinned extended maximum-likelihood fit to
the three variables $\mbc$, $\DeltaE$, and $\nbprim$. We define a probability
density function (PDF) for each event category, i.e., signal, $\qqbar$, and
$\BB$ backgrounds:
\begin{equation}
{\cal P}_{j}^{i}\equiv\tfrac{1}{2}(1-q_{i}\ACPj){\cal P}_j(M_{\rm bc}^{i}){\cal P}_j(\DeltaE^{\,i}){\cal P}_j(\nb'^{\,i}),
\label{eq:pdf}
\end{equation}
where $i$ denotes the event index, $q_{i}$ is the charge of the $B$ candidate 
($q_{i}=\pm 1$ for $\Bpm$), and ${\cal P}_{j}$ and $\ACPj$ are the PDF and $\CP$
asymmetry, respectively, for the event category $j$. The latter is defined as
\begin{eqnarray}
 \ACP =\dfrac{N_{\Bm}-N_{\Bp}}{N_{\Bm}+N_{\Bp}}, 
\end{eqnarray}
where $N_{\Bp}$ ($N_{\Bm}$) is the number of $\Bp$ ($\Bm$) events. We find equal
detection efficiencies for the $\Bp$ $(12.3\pm 0.1\%)$ and $\Bm$ $(12.4\pm 0.1\%)$
decays. For neutral $B$ decays, we replace the factor $\tfrac{1}{2}(1-q_{i}\ACPj)$
by $1$ in Eq.~(\ref{eq:pdf}). We also do not perform a $\CP$-violation study in this
case, since we would need to tag the recoiling $B$ candidate for that, causing
further loss in efficiency on top of the small signal yield. As the correlations
among $\mbc$, $\DeltaE$, and $\nbprim$ are found to be small ($\lesssim 5\%$), the
product of three individual PDFs is a good approximation for the total PDF. The
extended likelihood function is
\begin{eqnarray}
{\cal L}=\dfrac{e^{-\sum_{j}n_{j}}}{N!}\prod_{i}\Big[\sum_{j} n_{j}{\cal P}_{j}^{i}\Big]\mbox{,}
\end{eqnarray}
where $n_{j}$ is the yield of event category $j$, and $N$ is the total number of
candidate events. From the fitted signal yield ($n_{\mathrm{sig}}$), we calculate
the branching fraction as
\begin{equation}
 \mathcal{B}(B\to\PHI\PHI K)=\dfrac{n_{\mathrm{sig}}}{\varepsilon N_{\BB}[\mathcal{B}(\PHI\to\Kp\Km)]^{2}}\mbox{,} 
\label{eq:bf}
\end{equation}
where $\varepsilon$ and $N_{\BB}$ are the detection efficiency and the number of $\BB$
events, respectively. In case of $\Bz\to\PHI\PHI\Kz$, we multiply the denominator by a
factor of $\tfrac{1}{2}$ to account for $\Kz\to\KS$, as well as by the subdecay branching
fraction ${\mathcal B}(\KS\to\pip\pim)$~\cite{PDG}.

As the expected yield of the nonpeaking $\BB$ background is small, and it is distributed
similarly to $\qqbar$ in $\mbc$ and $\DeltaE$, we merge $\qqbar$ and $\BB$ backgrounds into
a single component. We find that the difference in the $\nbprim$ distribution between the
two backgrounds contributes a negligible systematic uncertainty. Table~\ref{tab:pdf-shape}
lists the PDF shapes used to model $\mbc$, $\DeltaE$, and $\nbprim$ distributions for
various event categories of $B\to\PHI\PHI K$ candidates. The yield and PDF shape parameters
of the combined background are floated in $\Bp\to\PHI\PHI\Kp$. For the neutral channel,
however, the background PDF shapes are fixed to their MC values after correcting for small
differences between data and simulation, as obtained from the charged decay. Similarly,
for the signal components, we fix the $\mbc$, $\DeltaE$, and $\nbprim$ shapes to MC values
and correct for small data-MC differences according to values obtained from a control sample
of $\Bp\to\Dsp\Dzb$ decays, where $\Dsp\to\PHI(\to\Kp\Km)\pi^{+}$ and $\Dzb\to\Kp\pi^{-}$. 

\begin{figure}[!htb]
\begin{center}
\includegraphics[width=.98\columnwidth]{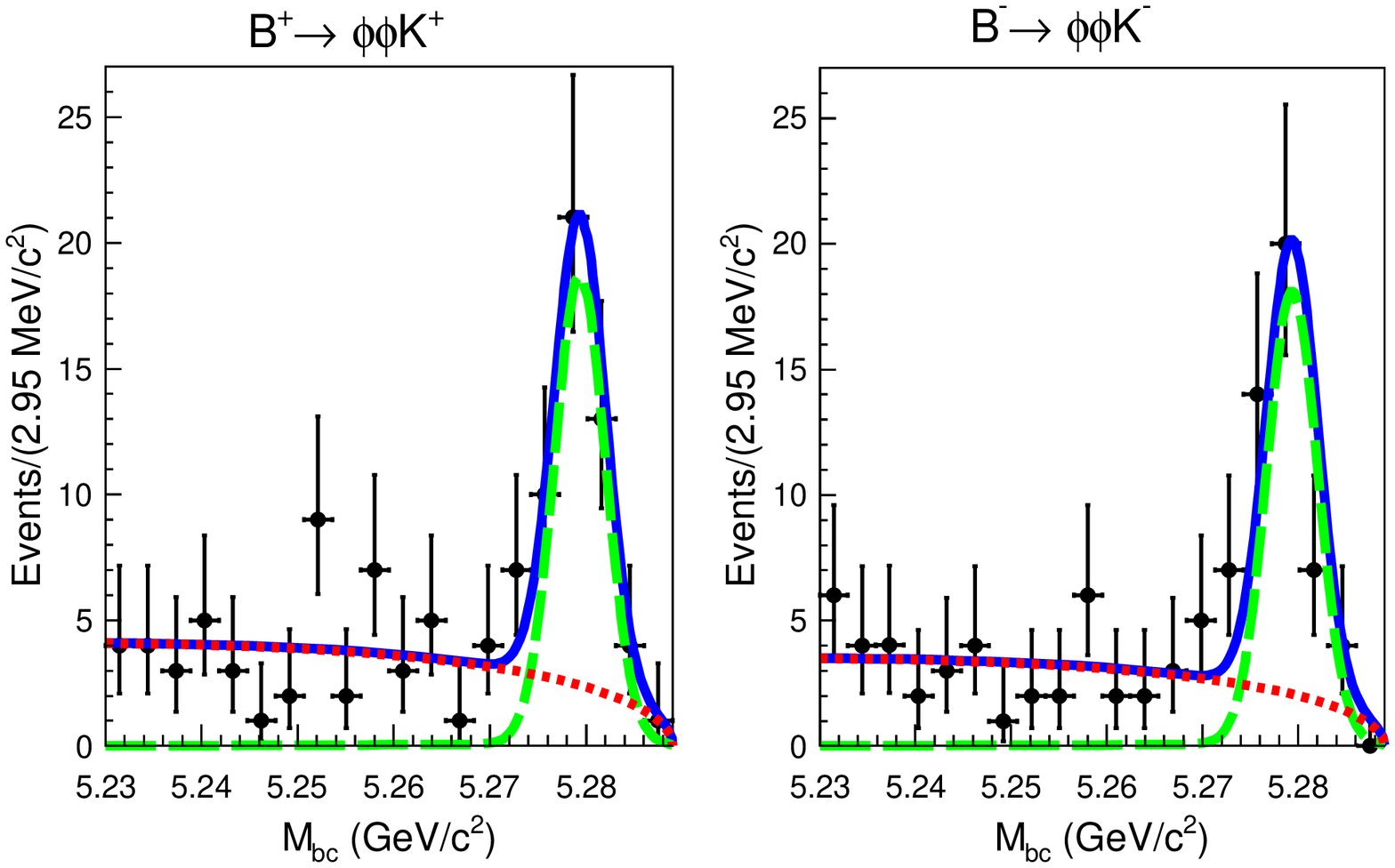}
\includegraphics[width=.98\columnwidth]{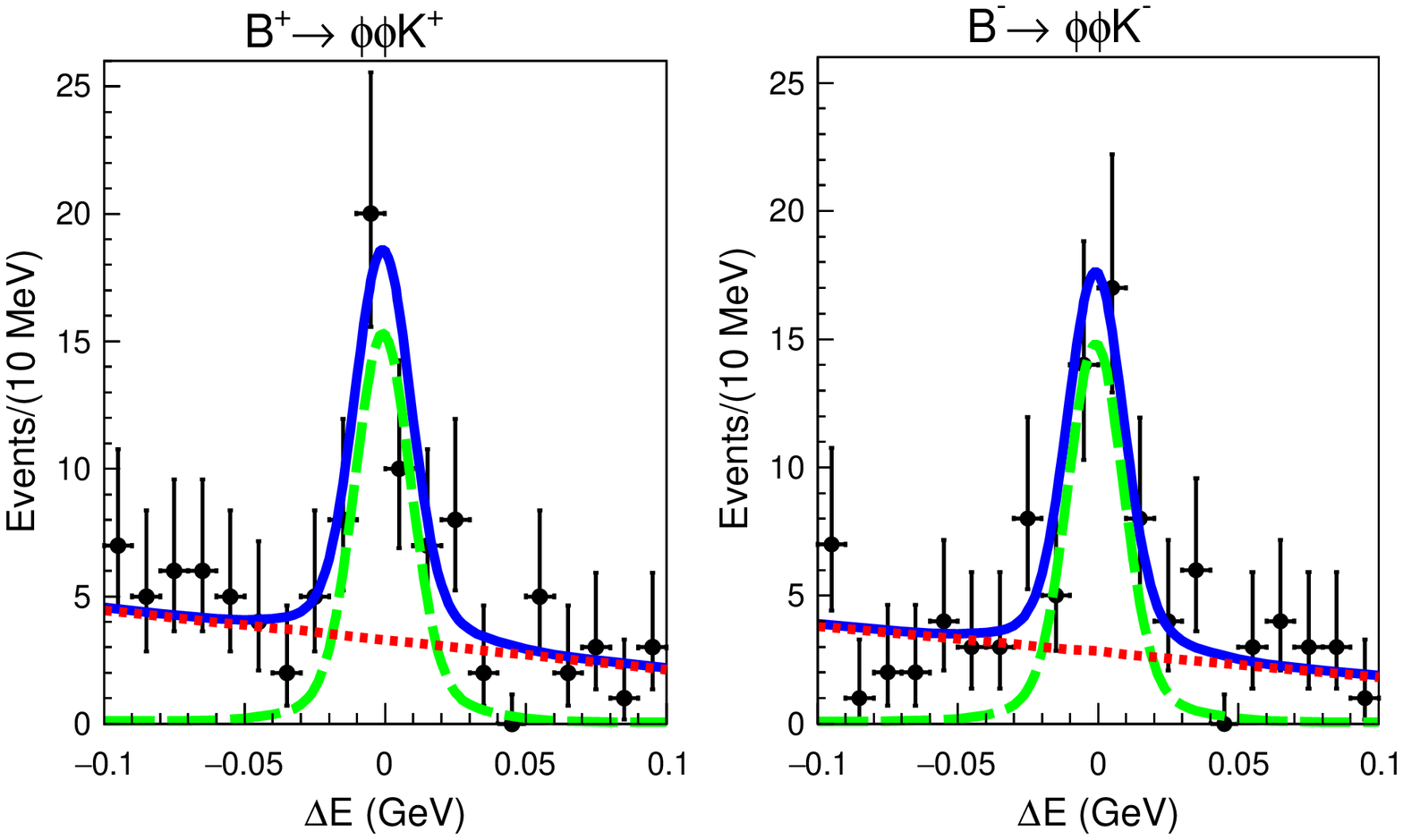}
\includegraphics[width=.98\columnwidth]{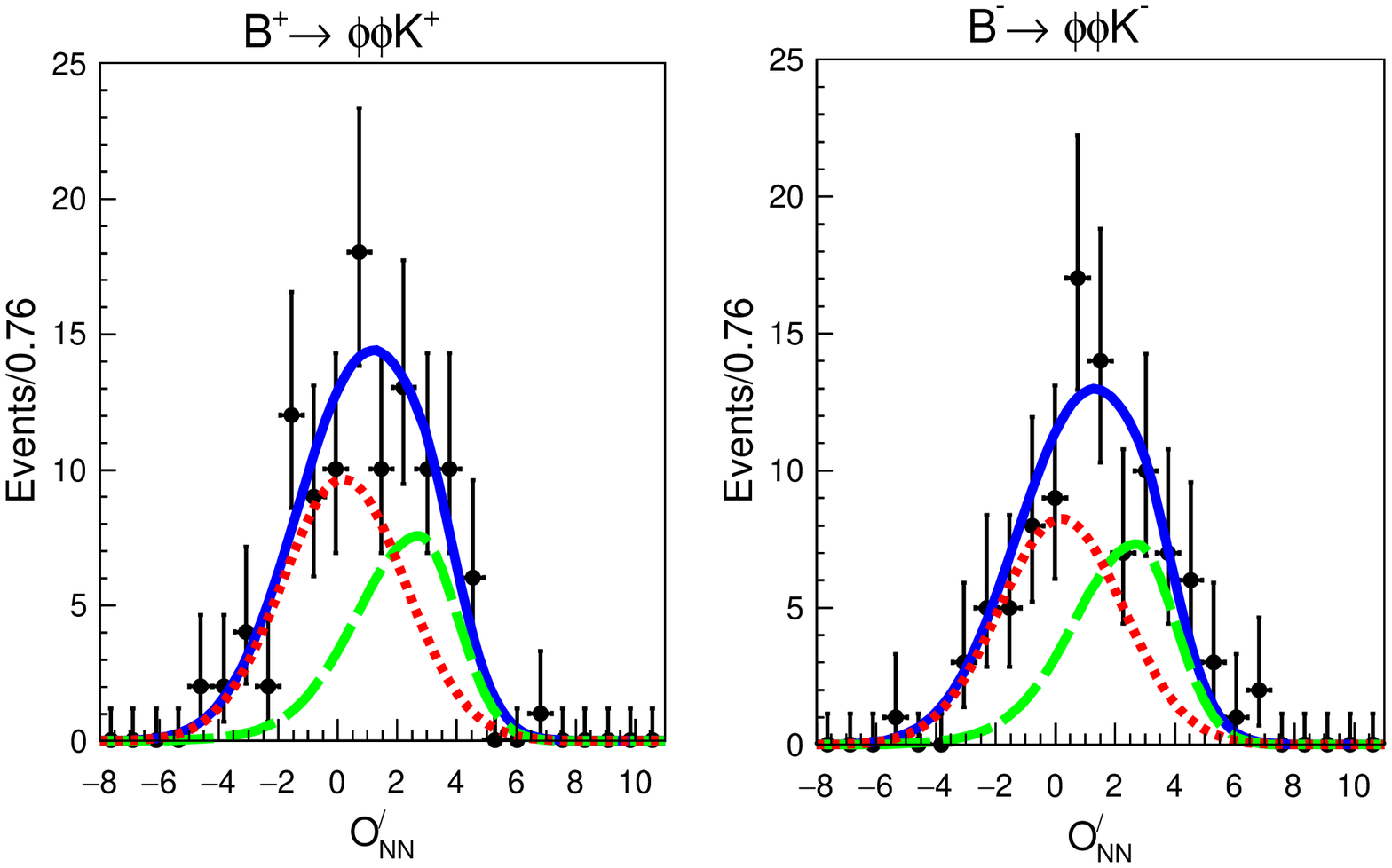}
\end{center}
\caption{Projections of $\Bpm\to\PHI\PHI\kpm$ candidate events onto (top) $\mbc$,
(middle) $\DeltaE$, and (bottom) $\nbprim$. Black points with error bars are the
data, solid blue curves are the total PDF, dashed green curves are the signal
component, and dotted red curves are the combined $\qqbar$ and $\BB$ background
components.}
\label{fig:3Da}
\end{figure}

\begin{figure}[!htb]
\begin{center}
\includegraphics[width=.98\columnwidth]{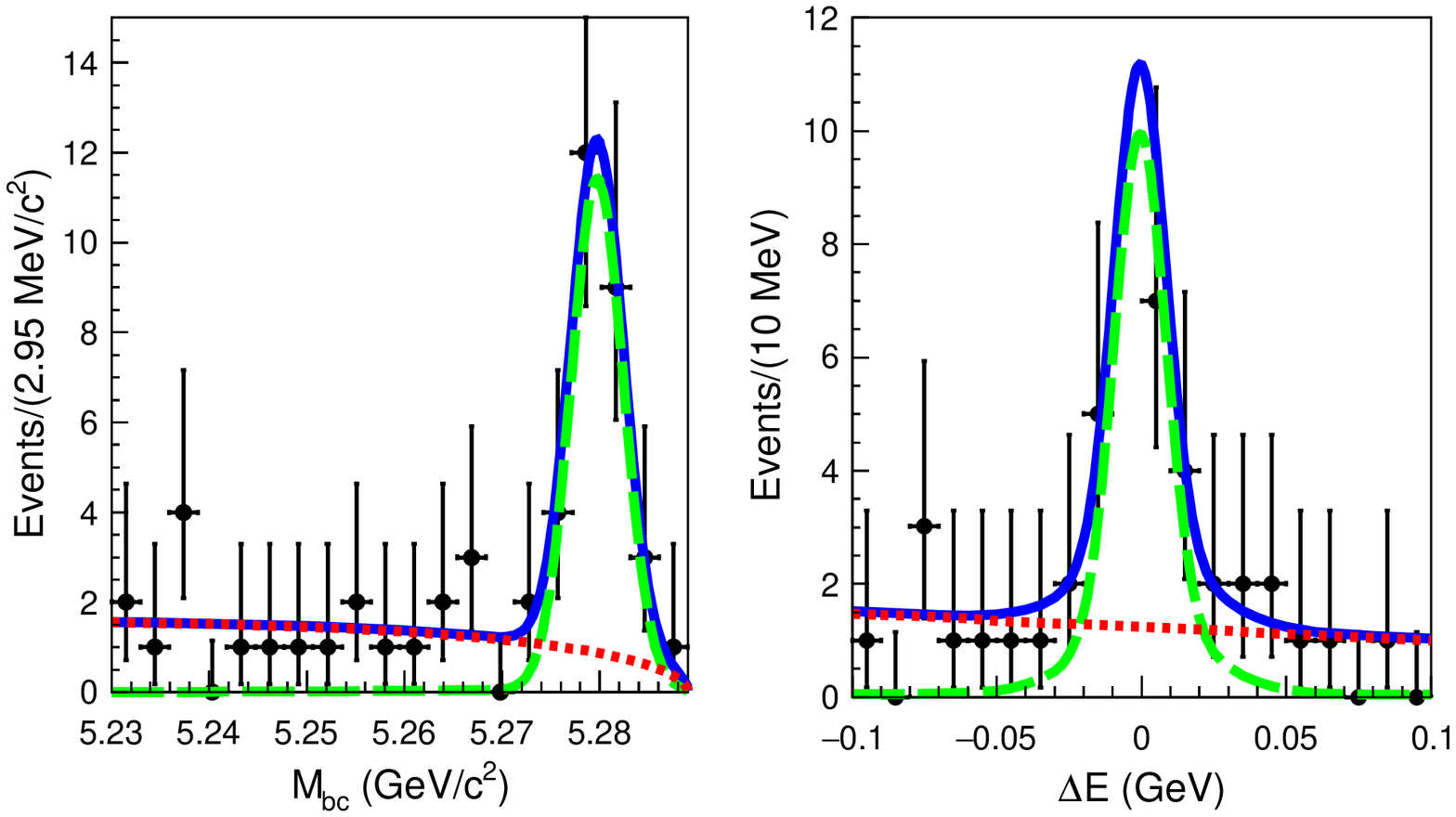}
\includegraphics[width=.98\columnwidth]{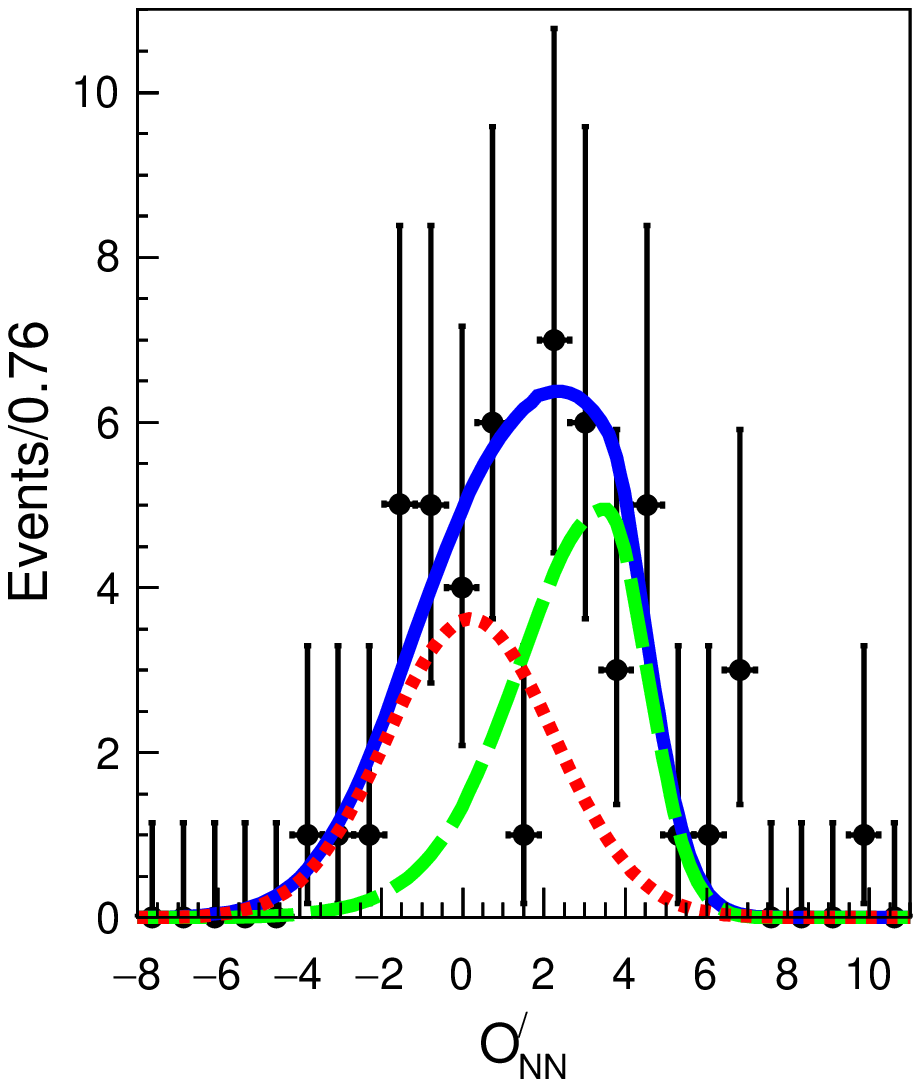}
\end{center}
\caption{Projections of $\Bz\to\PHI\PHI\Kz$ candidate events onto (top left)
$\mbc$, (top right) $\DeltaE$, and (bottom) $\nbprim$. The legends of the
plots are defined in the same manner as in Fig.~\ref{fig:3Da}.}
\label{fig:3Db}
\end{figure}

\begin{figure}[!htb]
\begin{center}
\includegraphics[width=.98\columnwidth]{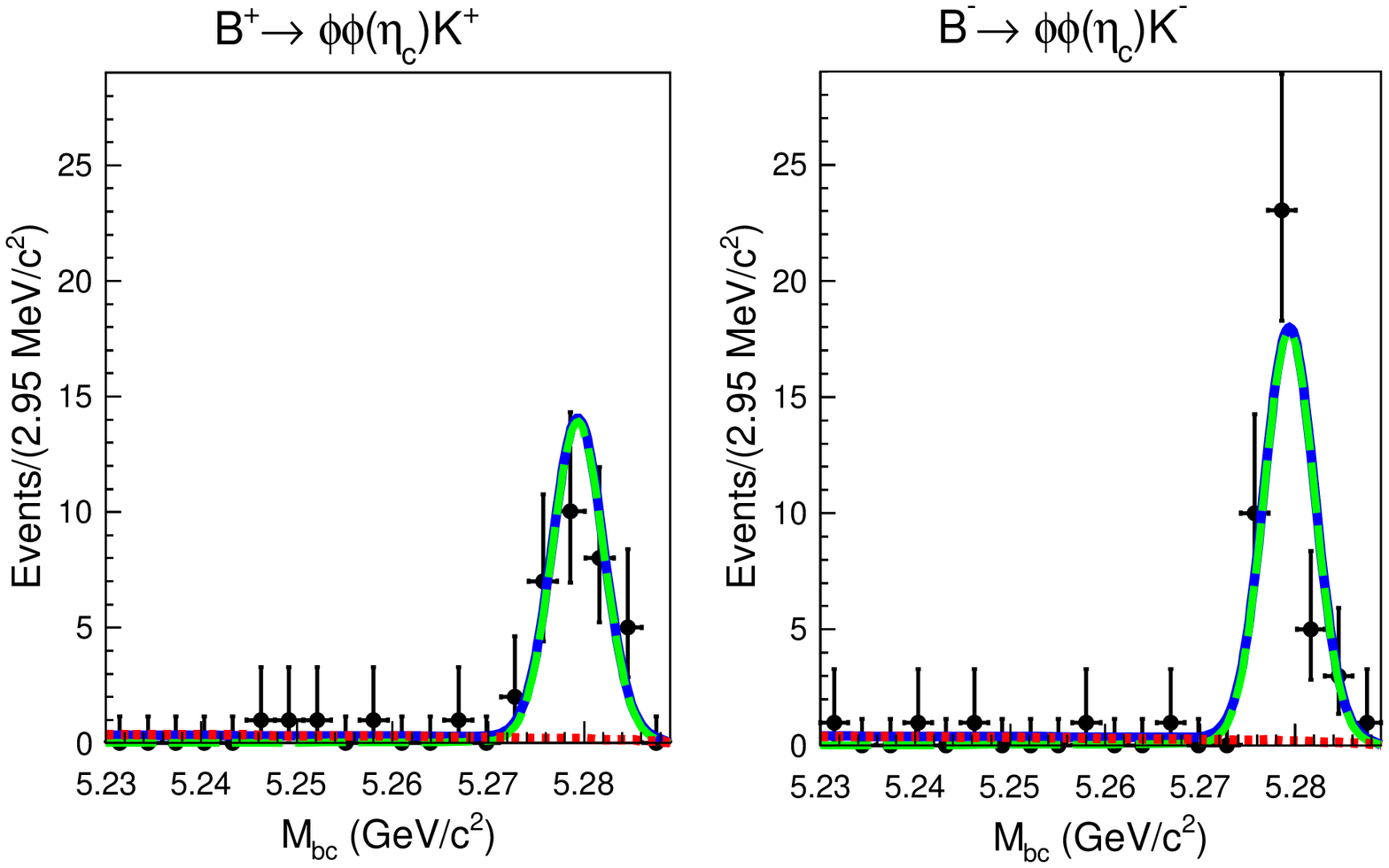}
\includegraphics[width=.98\columnwidth]{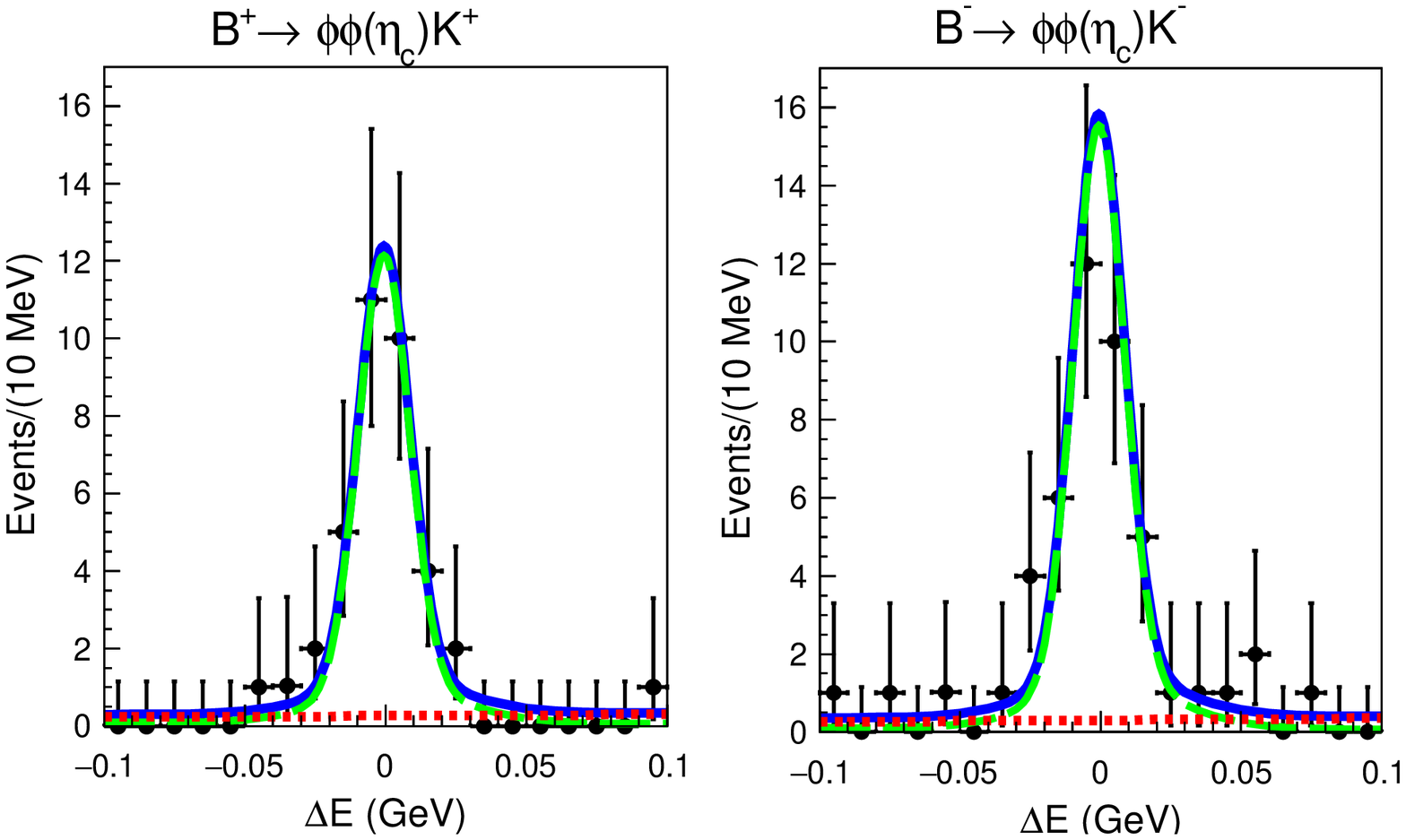}
\end{center}
\caption{Projections of $\Bpm\to\PHI\PHI\kpm$ candidate events within the
$\ETAc$ region onto (top) $\mbc$ and (bottom) $\DeltaE$. The legends of
the plots are defined in the same manner as in Fig.~\ref{fig:3Da}.}
\label{fig:3Dc}
\end{figure}

We apply the above 3D fit to $\Bp\to\PHI\PHI\Kp$ and $\Bz\to\PHI\PHI\Kz$
candidate events to determine the signal yield (and $\ACP$ in the first case).
Figures~\ref{fig:3Da} and \ref{fig:3Db} show $\mbc$, $\DeltaE$, and $\nbprim$
projections of the fits. The fit results are listed in Table~\ref{tab:BFs}. We
find signal yields of $85.0^{\,+\,10.2}_{\,-\,9.5}$ for $\Bp\to\PHI\PHI\Kp$
and $26.5^{\,+\,5.8}_{\,-\,5.1}$ for $\Bz\to\PHI\PHI\Kz$, and an $\ACP$ value of
$-0.02\pm0.11$ for the first case. We also apply the 3D fit to $\Bp\to\PHI\PHI
\Kp$ candidate events with $m_{\phi\phi}$ within the $\ETAc$ region to calculate
the signal yield and $\ACP$ value. The corresponding $\mbc$ and $\DeltaE$
projections are shown in Fig.~\ref{fig:3Dc}, with the fit results listed in
Table~\ref{tab:BFs}. We obtain a signal yield of $73.2^{\,+\,9.0}_{\,-\,8.3}$
and an $\ACP$ value of $+0.12\pm0.12$ in the $\ETAc$ region. The signal significance
is calculated as $\sqrt{-2\log({\cal L}_0/{\cal L}_{\rm max})}$, where ${\cal L}_0$
and ${\cal L}_{\rm max}$ are the likelihood values with the signal yield
fixed to zero and for the nominal fit, respectively. We include systematic
uncertainties that impact only the signal yield into the likelihood curve
via a Gaussian convolution before calculating the final significance.

\begin{table}[!htb]
\centering
\caption{List of PDFs used to model the $\mbc$, $\DeltaE$, and, $\nbprim$ distributions
for various event categories for $B\to\PHI\PHI K$. The notation G, AG, 2G, ARG, and Poly1
denote Gaussian, asymmetric Gaussian, sum of two Gaussians, ARGUS~\cite{argus} function,
and first-order polynomial, respectively.}
\label{tab:pdf-shape}
\begin{tabular}{lcccccc}
\hline\hline
Event category & & $\mbc$ & & $\DeltaE$ & & $\nbprim$ \\
\hline
Signal & & G+ARG & & 2G+Poly1 & & G+AG \\
$\qqbar+\BB$ & & ARG & & Poly1 & & G \\
\hline
\end{tabular}
\end{table}  

\begin{figure}[!htb]
\begin{center}
\includegraphics[width=.99\columnwidth]{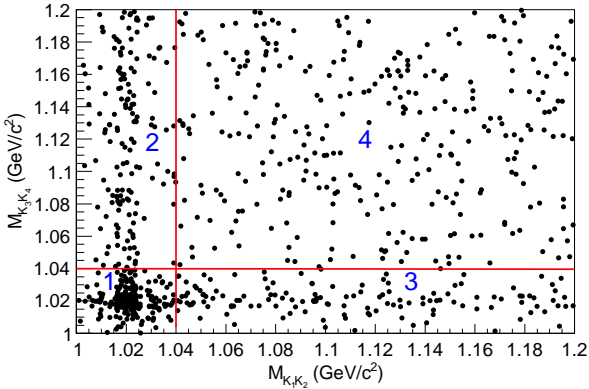}
\end{center}
\caption{Distribution of data events in the $M_{K_{1}K_{2}}$ vs $M_{K_{3}K_{4}}$
plane which shows the $M_{KK}$ signal region (region 1) and two sidebands SB1
(region 2 and 3) and SB2 (region 4).}
\label{fig:MK1K2_MK3K4}
\end{figure}

To estimate the contribution of $B\to\phi KKK$ and $B\to KKKKK$ decays in the
$M_{KK}$ signal region (SR), we repeat the 3D fit in the following two sidebands:
SB1 is denoted by the sum of ($M_{K_{1}K{2}}\in[1.04,1.2]\gevcc$ and $M_{K_{3}K{4}}
\in[1.0,1.04]\gevcc$) and ($M_{K_{1}K{2}}\in [1.0,1.04]\gevcc$ and $M_{K_{3}K{4}}\in
[1.04,1.2]\gevcc$), and SB2 is denoted by $M_{K_{1}K{2}}\in [1.04,1.2]\gevcc$ and
$M_{K_{3}K{4}}\in[1.04,1.2]\gevcc$. In Fig.~\ref{fig:MK1K2_MK3K4} we plot the
distribution of data events in the $M_{K_{1}K_{2}}$ vs $M_{K_{3}K_{4}}$ plane
showing SR, SB1, and SB2. The resonant $B\to\PHI\PHI K$ yield in SR is obtained
by solving the following three linear equations:
\begin{eqnarray}
N_{0}=n_{s}+r_{a0}\times n_{a}+r_{b0}\times n_{b},\\
N_{1}=r_{s1}\times n_{s}+n_{a}+r_{b1}\times n_{b},\\
N_{2}=r_{s2}\times n_{s}+r_{a2}\times n_{a}+n_{b},
\end{eqnarray}
where $N_{0}$, $N_{1}$, and $N_{2}$ are the yields obtained in SR, SB1, and SB2,
respectively; $n_{s}$, $n_{a}$, and $n_{b}$ are the $B\to\phi\phi K$ yield in SR,
$B\to\phi KKK$ yield in SB1, and $B\to KKKKK$ yield in SB2, respectively. Lastly,
$r_{s1}$ and $r_{s2}$ are the ratios of $B\to\phi\phi K$ yields in SB1 and SB2
to that in SR; $r_{a0}$ and $r_{a2}$ are the ratios of $B\to\phi KKK$ yields
in SR and SB2 to that in SB1; and $r_{b0}$ and $r_{b1}$ are the ratios of $B\to
KKKKK$ yields in SR and SB1 to that in SB2. All these ratios are obtained from
an MC study. We obtain the resonant $B\to\phi\phi K$ yield in SR ($n_{s}$) as
$81.8^{\,+\,10.1}_{\,-\,9.4}$ and $23.7^{\,+\,5.7}_{\,-\,5.0}$ for the charged
and neutral mode, respectively. These $n_{s}$ values are used in the branching
fraction calculation of Eq.~(\ref{eq:bf}).

\begin{table}[htb]
\centering
\caption{Number of candidate events ($n_{\rm cand}$), detection efficiency
($\varepsilon$), total and resonant signal yield ($n_{\rm sig}$), significance,
branching fraction ($\cal B$) and $\CP$ asymmetry ($\ACP$) obtained from a fit
to data for $B\to\PHI\PHI K$ decays below and within the $\ETAc$ region. Quoted
uncertainties are statistical only, and significances defined in the text are
given in terms of standard deviations.}
\label{tab:BFs}
\begin{tabular}{lccc}
\hline\hline
       & $\Bp\to\PHI\PHI\Kp$ & $\Bz\to\PHI\PHI\Kz$ & $\Bp\to\PHI\PHI(\ETAc)\Kp$ \\
\hline
$n_{\rm cand}$ & $207$  & $51$ & $84$    \\
$\varepsilon\,($\%$)$  & $12.4$ & $12.0$ & $15.4$    \\
Total $n_{\rm sig}$ & $85.0^{\,+\,10.2}_{\,-\,9.5}$ & $26.5^{\,+\,5.8}_{\,-\,5.1}$ & $73.2^{\,+\,9.0}_{\,-\,8.3}$\\
Significance & $14.9$ & $7.2$  &  $16.7$   \\
Resonant $n_{\rm sig}$ & $81.8^{\,+\,10.1}_{\,-\,9.4}$ & $23.7^{\,+\,5.7}_{\,-\,5.0}$ & -- \\
${\cal B}\,(10^{-6})$ & $3.43^{\,+\,0.48}_{\,-\,0.46}$ & $3.02^{\,+\,0.75}_{\,-\,0.66}$ & -- \\
$\ACP$ & $-0.02\pm0.11$ & -- & $+0.12\pm0.12$ \\
\hline
\end{tabular}
\end{table}

\begin{figure}[!htb]
\begin{center}
\includegraphics[width=.99\columnwidth]{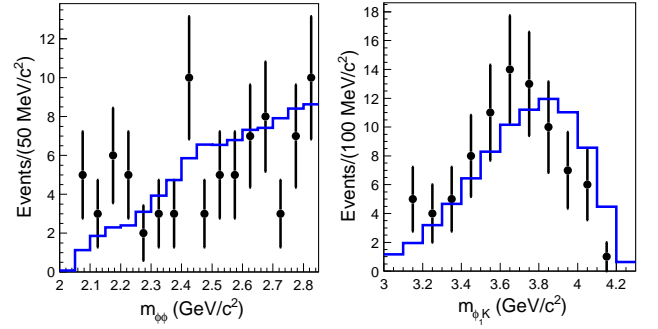} 
\end{center}
\caption{Background-subtracted signal yield as a function of $m_{\PHI\PHI}$ (left) and
$m_{\PHI_{1}K}$ (right) for $\Bpm\to\PHI\PHI\kpm$. Black points with error bars are data
and solid blue histograms denote the expectation from a phase-space MC sample.}
\label{fig:3Dd}
\end{figure}

The background-subtracted distributions~\cite{splot} of $m_{\PHI\PHI}$ and $m_{\PHI K}$
obtained for $\Bpm\to\PHI\PHI\kpm$ below the $\ETAc$ threshold are shown in Fig.~\ref{fig:3Dd}.
These are broadly compatible with the predictions of a three-body phase space MC sample.
In particular, we do not find any enhancement in the $m_{\PHI\PHI}$ spectrum, including
the $2.3\gevcc$ region~\cite{Ref:Chua} where a glueball and $X(2350)$ candidates are
predicted.

Systematic uncertainties in the branching fraction are listed in Table~\ref{tab:syst2}. The
uncertainties due to PDF shapes are estimated by varying all the fixed shape parameters by
their errors. In particular, for fixed signal shape parameters, we vary the data-MC corrections
by their uncertainties as determined using the control sample of $\Bp\to\Dsp\Dzb$ decays.
Potential fit bias is checked by performing an ensemble test comprising $1000$ pseudo-experiments,
where signal is taken from the corresponding MC sample, and the PDF shapes are used to generate
background events. We obtain a Gaussian normalized residual distribution of unit width, and add
its mean and uncertainty in width in quadrature to calculate the systematic error. Uncertainty
due to continuum suppression is obtained with the $\Bp\to\Dsp\Dzb$ control sample by comparing,
between data and simulation, fit results obtained with and without the $\nb$ requirement. A
$\Dstarp\to\Dz(\Km\pip)\pip$ control sample is used to determine the systematic uncertainty due
to the ${\cal R}_{K/\pi}$ requirement. We use partially reconstructed $\Dstarp\to\Dz(\KS\pip\pim)
\pip$ decays to assign the systematic uncertainty due to charged-track reconstruction ($0.35\%$ per
track). The uncertainty due to $\KS$ reconstruction is estimated from $\Dz\to\KS\KS$ decays~\cite{Ks}.
We estimate the uncertainty due to efficiency variation across the Dalitz plot by weighting
phase-space-generated signal MC events according to the measured distribution in data (Fig.~\ref{fig:3Dd})
and taking the difference between the weighted and nominal efficiency. The total systematic
uncertainty is obtained by adding all the above contributions in quadrature.

\begin{table}[!htb]
\centering
\caption{Systematic uncertainties (in $\%$) in the branching fractions. Values listed
in the top three rows impact the signal yield and are included in the calculation
of signal significance.}
\label{tab:syst2}
\begin{tabular}{lccc}
\hline\hline
Source & $\Bpm\to\PHI\PHI\kpm$ & $\Bz\to\PHI\PHI\Kz$ \\
\hline
Signal PDF & $^{\,+\,1.5}_{\,-\,1.7}$ & $^{\,+\,1.3}_{\,-\,1.9}$ \\
Background PDF & -- & $^{\,+\,3.0}_{\,-\,1.9}$ \\
Fit bias & $\pm 1.7$ & $\pm 2.0$ \\
Efficiency variation & $\pm 2.1$ & $\pm 2.1$ \\
${\cal R}_{K/\pi}$ requirement & $\pm 5.2$ & $\pm 4.3$ \\
$\qqbar$ suppression & $\pm 0.5$ & $\pm 0.5$ \\
Track reconstruction & $\pm 1.8$ & $\pm 1.4$ \\
$\KS$ reconstruction & -- & $\pm 0.9$ \\
Number of $\BB$ events & $\pm 1.4$ & $\pm 1.4$ \\
\hline
Total & $\pm 6.5$ & $^{\,+\,6.5}_{\,-\,6.3}$ \\
\hline
\end{tabular}
\end{table}  

We consider two possible sources of systematic uncertainties contributing to $\ACP$, as listed
in Table~\ref{tab:syst3}. The first is due to the intrinsic detector bias on charged kaon detection
and is estimated using $\Dsp\to\phi\pi^{+}$ and $\Dz\to K^{-}\pi^{+}$ decays~\cite{det-bias}.
The second arises due to the potential variation of the PDF shapes. We calculate its contribution
by following a procedure similar to that used in estimating the PDF shape uncertainties in the
branching fractions.

\begin{table}[!htb]
\caption{Systematic uncertainties in $\ACP$.}
\label{tab:syst3}
\begin{tabular}{lccc}
\hline\hline
Source & & \multicolumn{2}{c}{$\Bpm\to\PHI\PHI\kpm$ $\Bpm\to\PHI\PHI(\ETAc)\kpm$} \\
\hline
Detection asymmetry & & $\pm 0.008$ & $\pm 0.008$ \\
Signal PDF shape & & $^{\,+\,0.002}_{\,-\,0.003}$ & $\pm 0.002$ \\
\hline
Total & & $\pm0.01$ & $\pm0.01$ \\
\hline
\end{tabular}
\end{table}

In summary, we have measured the branching fractions and $\CP$-violation asymmetries in 
$B\to\PHI\PHI K$ decays based on the full $\Upsilon(4S)$ data sample of $772\times 10^{6}$
$\BB$ events collected by the Belle detector at the KEKB asymmetric-energy $\EP\EM$ collider.
We obtain the branching fraction and $\CP$ asymmetry for $\Bpm\to\PHI\PHI\kpm$ below the
$\ETAc$ threshold $(m_{\PHI\PHI}<2.85\gevcc)$ as
\begin{eqnarray}
(3.43^{\,+\,0.48}_{\,-\,0.46}\pm 0.22)\times 10^{-6}
\end{eqnarray}
and
\begin{eqnarray}
-0.02\pm 0.11\pm 0.01,
\end{eqnarray}
respectively. We also report the $\CP$-violation asymmetry for $\Bpm\to\PHI\PHI\kpm$
in the $\ETAc$ region ($m_{\PHI\PHI}\in [2.94,3.02]\gevcc$) to be
\begin{eqnarray}
+0.12\pm0.12\pm0.01,
\end{eqnarray}
consistent with no $\CP$ violation. The obtained value of the branching fraction
of $\Bpm\to\PHI\PHI\kpm$ decay is consistent with and supersedes our previous
result~\cite{Ref:Huang}. The measured branching fraction for $\Bz\to\PHI\PHI\Kz$
below the $\ETAc$ threshold is
\begin{eqnarray}
(3.02^{\,+\,0.75}_{\,-\,0.66}\pm 0.20)\times 10^{-6}.
\end{eqnarray}
We find no evidence for glueball production in these decays.

SM acknowledges fruitful discussions with S. Mahapatra (Utkal University).
We thank the KEKB group for excellent operation of the accelerator; the KEK cryogenics
group for efficient solenoid operations; and the KEK computer group, the NII, and 
PNNL/EMSL for valuable computing and SINET5 network support. We acknowledge support
from MEXT, JSPS and Nagoya's TLPRC (Japan); ARC (Australia); FWF (Austria); NSFC
and CCEPP (China); MSMT (Czechia); CZF, DFG, EXC153, and VS (Germany); DAE, Project
Identification No. RTI 4002, and DST (India); INFN (Italy); MOE, MSIP, NRF, RSRI,
FLRFAS project, GSDC of KISTI and KREONET/GLORIAD (Korea); MNiSW and NCN (Poland);
MSHE, Agreement 14.W03.31.0026 (Russia); University of Tabuk (Saudi Arabia);
ARRS (Slovenia); IKERBASQUE (Spain); SNSF (Switzerland); MOE and MOST (Taiwan);
and DOE and NSF (USA).


\begin{thebibliography}{99}
\bibitem{Ref:Inclusion}
 Inclusion of charge-conjugate reactions is implied unless stated otherwise.
\end{thebibliography}
\end{document}